\documentclass[12pt]{article}

\usepackage{graphicx}
\usepackage{color}
\usepackage{bm}
\def\beq{\begin{equation}}
\def\eeq{\end{equation}}
\def\beqa{\begin{eqnarray}}
\def\eeqa{\end{eqnarray}}
\def\bet{\begin{tabular}}
\def\eet{\end{tabular}}

\def\dbar{\Delta} %\def\dbar{\bar\Delta}
\newcommand{\ex}[1]{{\rm e}^{#1}} \def\ii{{\rm i}}

\newcommand{\sect}[1]{\setcounter{equation}{0}\section{#1}}

\def\one{{\hbox{ 1\kern-.8mm l}}}

\usepackage[centertags]{amsmath}

\usepackage{amsfonts} 
\usepackage{amssymb} 
\usepackage{amsthm}
\newcommand{\be}{\begin{equation}}
\newcommand{\ee}{\end{equation}}
\newcommand{\bea}{\begin{eqnarray}}
\newcommand{\eea}{\end{eqnarray}}
\newcommand{\bean}{\begin{eqnarray*}}
\newcommand{\eean}{\end{eqnarray*}}
\newcommand{\nn}{\nonumber}
\newcommand{\ft}[2]{{\textstyle {\frac{#1}{#2}} }}

\begin{document}

\begin{titlepage}

\renewcommand{\thefootnote}{\alph{footnote}}

\begin{flushright}
{QMUL-PH-09-27}
\\
{ROM2F/2009/27}
\end{flushright}

\vspace{0.6cm}

\begin{center}
{\Large \bf D1D5 microstate geometries from string amplitudes} \\ 

\vskip 0.8cm

{\bf Stefano Giusto}\footnote{stefano.giusto@cea.fr}\\
{\sl Laboratoire de Physique Th\'eorique et Hautes Energies}\\
{\sl Universit\'e Pierre et Marie Curie - Paris 6}\\
{\sl 4 Place Jussieu, 75252 Paris cedex 05, France}

\vskip .3cm

{\bf Jose F. Morales}\footnote{morales@roma2.infn.it}\\
{\sl INFN, Universita di Roma "Tor Vergata"}\\
{\sl Via delle ricerca scientifica 1, 00183, Roma, Italia}\\

\vskip .3cm

{\bf Rodolfo Russo}\footnote{r.russo@qmul.ac.uk}\\
{\sl 
Centre for Research in String Theory \\ 
Department of Physics, Queen Mary University of London\\
Mile End Road, London, E1 4NS,
United Kingdom}\\

\vskip 1.2cm

\end{center}

\begin{abstract}

We reproduce the asymptotic expansion of the D1D5 microstate
geometries by computing the emission amplitudes of
closed string states from disks with mixed D1D5 boundary conditions.
Thus we provide a direct link between the supergravity and D-brane
descriptions of the D1D5 microstates at non-zero string
coupling. Microscopically, the profile functions characterizing the
microstate solutions are encoded in the choice of a condensate for the
twisted open string states connecting D1 and D5 branes.

\end{abstract}

\vfill

\end{titlepage}

\renewcommand{\thefootnote}{\arabic{footnote}} 	
\setcounter{footnote}{0}

\sect{Introduction}

The gravitational description of black hole microstates remains one of
the fundamental and most debated problems, more than ten years after
the seminal works \cite{Sen:1995in, Strominger:1996sh} .  While these
works have shown that, at weak coupling, black hole microstates can be
described and counted in terms of D-branes, their description when the
gravitational coupling becomes finite is still considered an unsettled
question. A possible answer to this question is provided by the
``fuzzball proposal'' (for reviews see \cite{Mathur:2005zp,
Bena:2007kg, Skenderis:2008qn, Balasubramanian:2008da}): in its
essence, it states that the `naive' geometry of black holes, in which
the horizon is locally empty space, is modified by quantum gravity
corrections up to scales of the horizon size. The proposal is
motivated by the explicit construction of classical geometries with
the same asymptotics than the black hole (or black string) solutions
but with no horizon.  The best understood example is that of the
2-charge BPS system in type IIB string theory, for which the
geometries dual to the whole family of microstates has been
constructed in \cite{Lunin:2001fv, Lunin:2001jy, Lunin:2002bj,
Lunin:2002iz, Taylor:2005db, Kanitscheider:2007wq}. The 2-charge
system represents a somewhat degenerate example of black hole, in that
its classical geometry has a singular horizon of zero size, and one
needs higher derivative corrections to smooth out the singularity and
produce a finite size horizon. It is thus of crucial importance to
extend the construction of the geometries dual to microstates to the
case of the 3-charge BPS black hole, which is likely to share the
properties of general (extremal) black holes. Though many advances
have been made towards this goal \cite{Lunin:2004uu, Giusto:2004id, Giusto:2004kj,
Bena:2005va, Berglund:2005vb, Bena:2006kb, Bena:2008nh}, a complete
construction of 3-charge microstates is still missing. Part of the
reason why this task has proved to be so challenging is that one lacks
a systematic procedure to construct the geometry generated by a
particular D-brane configuration.

To help close this gap, in this paper we provide the relation between
the stringy description of the black hole microstates in terms of
D-branes and the corresponding geometries of the gravitational
description. In principle, once a D-brane configuration is fully
understood at zero string coupling ($g_s=0$), it is sufficient to
switch on a non-zero coupling in order to derive the gravitational
backreaction of the system. When $g_s\ll 1$ it is possible to use the
conformal field theory description of D-branes to compute the
corresponding gravitational profile perturbatively. In practice, one
needs to compute disc amplitudes with the insertion of a closed string
vertex: the boundary conditions on the disc should carry all the
information of the D-brane configuration under analysis, while the
closed string emitted represents the gravitational backreaction. For
the half-BPS geometries the situation is quite simple and this
mechanism has been checked explicitly
in~\cite{DiVecchia:1997pr}. However this idea should work for any
supergravity state that admits a description in terms of D-branes and,
for instance, in~\cite{Bertolini:2000jy} this approach was used to
study a non-BPS configuration.

The situation we want to study is that of D-brane {\em bound states}
where the single constituents of the system cannot be freely
separated. The first non-trivial case is represented by the 2-charge
systems in type IIB string theory.
There are various dual descriptions of these configurations. For
instance the geometries corresponding to each microstate of this
system have been derived~\cite{Dabholkar:1995nc, Callan:1995hn,
Lunin:2001fv} by using the description in terms of a fundamental
string with a wave carrying a left (or right) moving momentum. These
solutions are specified by a {\it profile function}, characterizing
the profile of the fundamental string.  Then by using a chain of
dualities these geometries can be re-interpreted as solutions
describing D1D5 bound states. Alternatively, these solutions can be
derived by solving the $\ft14$-BPS killing spinor equations of
supergravity and can be thought as ``bubblings" of the naive D1D5
geometry \cite{Martelli:2004xq}.

In this paper we address the study of the microstate geometries from
the string theory perspective.  In particular we will show how the
leading asymptotics of the microstate geometries are reproduced by
string amplitudes describing the emissions of closed string states
from D1D5 disks.  In principle we could start from the description of
a D-string with a wave~\cite{Bachas:2002jg}, derive the gravitational
backreaction by using the boundary state for these
D-branes~\cite{Hikida:2003bq} and then rewrite the result in the D1D5
duality frame. However, since our aim is to provide a direct link
between the D-brane construction and the corresponding geometries, we
will focus directly on the analysis of the D1D5 bound states. From the
conformal field theory point of view these bound states are described
by open string amplitudes with mixed boundary conditions. Mixed disks
dual to the ones considered here were studied in~\cite{Billo:2002hm},
where a direct link between the open string emission and the gauge
theory instantons was found. In this paper we will study the closed
string emission from disks that have half of their boundary along the
D1-branes and the other half along the D5-branes.  We see that these
simple amplitudes are sufficient to reproduce the first corrections
that distinguish the microstate geometries from the naive
superposition of D1 and D5-branes.  These corrections display some of
the fundamental properties of the fuzzball geometries, such as the
fact that the non-trivial states carry some angular momentum that
breaks the rotational invariance in the non-compact directions, and
the presence of a Kaluza-Klein monopole dipole charge, which is
ultimately responsible for the regularity of the geometries in their
core. In the string description they can be associated to condensates
of twisted open string states stretched between the D1 and D5 branes.
The open string condensate, as we will see, encodes the information
about the profile function characterizing the microstate supergravity
solution.

The paper is organized as follows. In Section~2 we review the
solutions for the D1D5 microstate geometries of IIB supergravity on
$\mathbb{R}^{4,1}\times S^1\times T^4$ and write explicitly the first
order corrections to the naive D1D5 superposition. In Section~3 we
introduce all CFT ingredients we will need to compute the mixed disc
amplitude mentioned above: we write the vertex operators for both the
closed and the open strings, and discuss the identification between
the holomorphic and antiholomorphic string coordinates induced by the
D-branes. In Section~4 we compute the emission of one closed string
state from a D1D5 disk and show that they reproduce the long distance
behavior of the microstate geometries reviewed in Section~2. Finally,
in the Conclusions, we discuss some possible generalisations and
applications of our results.

\sect{Review of D1D5 geometries}

Let us focus on type IIB string theory on $\mathbb{R}^{4,1}\times
S^1\times T^4$.  If one wraps $n_1$ D1 branes on $S^1$ and $n_5$ D5
branes on $S^1\times T^4$, one obtains a system that preserves $1/4$
of the 32 supersymmetries of type IIB strings and has, in the large
$n_1 n_5$ limit, $\exp{(\sqrt{2}\pi \sqrt{n_1 n_5})}$ states. The
supergravity description of all these states was found
in~\cite{Lunin:2001fv, Lunin:2001jy, Lunin:2002bj,Lunin:2002iz,
Taylor:2005db,Kanitscheider:2007wq}, exploiting the fact that the D1D5
system is U-dual to a fundamental string on $\mathbb{R}^{4,1}\times
S'^1\times T'^4$ wrapped $n_5$ times on $S'^1$ and carrying $n_1$
units of (left moving) momentum.  The states of the fundamental string
are described, in the semi-classical limit, by a curve in the space
transverse to the string, $\mathbb{R}^4\times T'^4$. We denote the
parametric representation of this curve by $f_A(v)$, with
$A=1,\ldots,8$, $v=t-y$ and $y$ the coordinate on $S'^1$ of radius
$R'$. The length of the multiply wound fundamental string is $L=2\pi
n_5 R'$, and the curve $f_A(v)$ is taken to have trivial winding along
the $T^4$ directions, so that $f_A(v+L)=f_A(v)$, for any $A$. It will
be convenient to distinguish the $\mathbb{R}^4$ directions, denoted by
the indices $i,j,\dots=1,\ldots,4$, from the $T^4$ directions, labeled
by $a,b,\ldots=5,\ldots,8$. Moreover the U-duality connecting the D1D5
and the fundamental string descriptions of the system requires to pick
a particular direction in the four dimensional torus. So the internal
components $f_a(v)$ of the curve describing the fundamental string are
not labeled by a vector index of the $T^4$ of the D1D5 description,
but by an index $\hat{a}$ running over the three self-dual
2-forms plus a scalar. In this notation the string profile is
represented, in the D1D5 duality frame, by the functions
\be
f_A(v) \equiv  (f_i(v),f_{\hat{a}}(v),f(v))\,.
\ee
We will choose the origin of our coordinates such as $\int_0^L f_A(v)=0$.

Along the time and the (compact) $y$ coordinate both the D1 and D5
branes have Neumann boundary conditions, while $x_i$ and $x_a$
parametrize $\mathbb{R}^4$ and $T^4$ where the D-branes have Dirichlet
and mixed Neumann/Dirichlet boundary conditions respectively.  

The
string frame metric ($ds^2$), dilaton ($\Phi$), B-field ($b$) and
p-form RR fields ($C^{(p)}$) of the generic D1D5 state are given by
\bea
ds^2 &\!\!=\!\!& {\hat H_1^{1/2}\over \tilde H_1 H_5^{1/2}} [-(dt-A_i dx_i)^2 + (dy+B_i dx_i)^2]+ (\hat H_1 H_5)^{1/2} dx_i dx_i + \Bigl({\hat H_1\over H_5}\Bigr)^{1/2} \!\! dx_a dx_a\,,\nonumber\\
e^{2\Phi}&\!\!=\!\!& {\hat H_1^2\over \tilde H_1 H_5}\,,\nonumber\\
b &\!\!=\!\!& - {\mathcal{A}\over \tilde H_1 H_5} (dt - A)\wedge (dy+B) + \mathcal{B} + {\mathcal{A}_{\hat{a}} \omega^{\hat{a}} \over H_5}\,,\nonumber\\
C^{(0)} &\!\!=\!\!& -{\mathcal{A}\over \hat H_1}\,,\nonumber\\
C^{(2)}&\!\!=\!\!& -dt\wedge dy +{\tilde H}_1^{-1}(dt-A)\wedge (dy+B)+C\,,\nonumber\\
C^{(4)}&\!\!=\!\!& - dt\wedge dy\wedge \mathcal{B} - {\mathcal{A}\over H_5 \tilde H_1}(dt-A)\wedge (dy+B)\wedge C - {\mathcal{A}\over H_5 \tilde H_1}(dt\wedge dy + C)\wedge A\wedge B\nonumber\\
&-& {\mathcal{A}_{\hat{a}}\over H_5} dt\wedge dy\wedge \omega^{\hat{a}} + (\mathcal{B}_{\hat{a}} + {\mathcal{A}_{\hat{a}}\over H_5} C)\wedge \omega^{\hat{a}} - {\mathcal{A}\over H_5} dx_{5} \wedge dx_{6}\wedge dx_{7}\wedge dx_8\,.
\label{generald1d5}
\eea
In our conventions the RR field strengths $F^{(p+1)}$ are defined as 
\be
F^{(p+1)}= d C^{(p)}-H^{(3)}\wedge C^{(p-2)}\,,
\ee 
with $H^{(3)}= d b$. The 5-form field strength is taken to be
self-dual\footnote{Our conventions for the Hodge star are
$(*a)_{i_1\ldots i_{d-p}}=\ft{1}{p!}\,\epsilon_{i_1\ldots i_{d-p}
j_1\ldots j_{p}} a^{j_1\ldots j_{p}}$, with $\epsilon_{01\ldots
d}=\sqrt{|g|}$.} $F^{(5)}=* F^{(5)}$. The various functions appearing
above are defined as follows
\bea\label{deff}
&&H_5 = 1+{Q_5\over L}\int_0^L {dv\over |x_i-f_i(v)|^2}\,,\quad H_1 = 1+ {Q_5\over L}\int_0^L {dv |\dot{f}_A(v)|^2\over |x_i-f_i(v)|^2}\,,\nonumber\\
&& %f_5 = H_5\,,
\quad {\hat H_1} = H_1 -{\mathcal{A}_{\hat{a}} \mathcal{A}_{\hat{a}}\over f_5}\,,\quad \tilde H_1 = H_1  -{\mathcal{A}_{\hat{a}} \mathcal{A}_{\hat{a}}+ \mathcal{A}\mathcal{A}\over f_5}\,,\nonumber\\
&&A_A = -{Q_5\over L}\int_0^L{dv \dot{f}_A(v)\over |x_i-f_i(v)|^2}\equiv (A_i,\mathcal{A}_{\hat{a}},\mathcal{A})\,,\quad A\equiv A_i dx_i\,,\\
&&d B = - *_4 dA\,,\quad dC = - *_4 d H_5\,,\quad d\mathcal{B}_{\hat{a}} = *_4 d \mathcal{A}_{\hat{a}} \,,\quad d\mathcal{B} = *_4 d\mathcal{A}\,, \nonumber
\eea
where $*_4$ denotes the Hodge dual
with respect to flat $\mathbb{R}^4$.  Finally the three 2-forms
$\omega^{\hat{a}}$ form a basis for the self-dual forms $\star_4
\omega^{\hat{a}} =\omega^{\hat{a}}$, where $\star_4$ is again defined
with respect to a flat metric, but now acts on the indices in the
$T^4$.  The length $L$ can be expressed in terms of D1D5 quantities as
\be 
L = 2\pi {Q_5\over R}\,,
\label{length}
\ee
with $R$ the radius of $S^1$ in the D1D5 frame. The D1 charge is given by
\be
Q_1= {Q_5\over L}\int_0^L  |\dot{f}_A(v)|^2\,,
\label{d1charge}
\ee
and the charges $Q_1$ and $Q_5$ are quantized as
\be
Q_1 = {(2\pi)^4 g \alpha'^3 n_1\over V_4}\,,\quad Q_5  = g \alpha' n_5\,,
\label{quant}
\ee
with $V_4$ the volume of $T^4$ and $n_1$, $n_5$ the numbers of D1, D5 branes.

\subsection{Asymptotic expansion of the geometry}

In the limit in which the backreaction of the D-branes on the geometry
is small, the D1D5 system can be described by the perturbative
dynamics of open strings stretched between the D-branes. This is the
description we will focus on in the next section. This description
should capture the large distance expansion of the geometries
(\ref{generald1d5}), which looks like a small perturbation around flat
space.

We focus on the terms that distinguish the microstate geometries
(\ref{generald1d5}) from the naive D1D5 geometry, which is the
singular, spherically symmetric, geometry with $f_A(v)=0$. These terms
first appear at order $1/r^3$ and are encoded in the functions 
$A_i$, $\mathcal{A}_{\hat{a}}$, $\mathcal{A}$ and in their duals
$B_i$, ${\cal B}_{\hat a ij}$, ${\cal B}_{ij}$. Their large distance
expansion is given by
\bea
A_i &\approx&-{Q_5\over L}\int_0^L\!\!\!dv \dot{f}_i \Bigl[{1\over r^2}+2 {x_j f_j\over r^4}\Bigr]= -2 Q_5 \hat f_{ij} {x_j\over r^4}\,,\quad \hat f_{ij} ={1\over L} \int_0^L\!\!\!dv \dot{f}_i f_j=-\hat f_{ji}\,,\label{r31}\\
\mathcal{A}_{\hat{a}} &\approx&-{Q_5\over L}\int_0^L \!\!\!dv\dot{f}_{\hat{a}} \Bigl[{1\over r^2}+2 {x_j f_j\over r^4}\Bigr]= -2 Q_5 \hat f_{\hat{a} j} {x_j\over r^4}\,,\quad \hat f_{\hat{a}j} ={1\over L} \int_0^L \!\!\!dv \dot{f}_{\hat{a}} f_j\,,\label{r32}\\
\mathcal{A} &\approx&-{Q_5\over L}\int_0^L \!\!\!dv\dot{f} \Bigl[{1\over r^2}+2 {x_j f_j\over r^4}\Bigr]= -2 Q_5 \hat f_{j} {x_j\over r^4}\,,\quad \hat f_{j} ={1\over L} \int_0^L \!\!\!dv \dot{f} f_j\,,\label{r33}
\eea
where we have used $\int_0^L \dot{f}_A=0$, and
\beq\label{r34}
B_i\approx -Q_5 \epsilon_{ijkl}\hat f_{kl} {x_j\over r^4}\;,\qquad
{\cal B}_{\hat a ij} \approx -2Q_5 \epsilon_{ijkl}
\hat f_{\hat{a} k} {x_l\over r^4}\;,\qquad
{\cal B}_{ij} \approx  -2Q_5 \epsilon_{ijkl} \hat f_{k} {x_l\over
  r^4}\,. 
\eeq

In this paper we consider D1D5 geometries that are invariant under
the $SO(4)$ acting along the ND directions (i.e. the $T^4$ coordinates).
Hence we will focus on the solutions that have $\mathcal{A}_{\hat{a}}=
{\cal B}_{\hat{a} ij}=0$. From (\ref{r31}-\ref{r34}) one finds that the
asymptotic form of such solutions in the large distance limit is
\bea
g_{ti}  &=& -{2 Q_5 x_j \hat f_{ij}\over r^4}~~~,
\quad\quad
g_{yi} =  - \epsilon_{ijkl}\, 
{Q_5 x_j \hat f_{kl}\over r^4 }~,\nn \\
b_{ty}  &=&  {2 Q_5 x_i \hat f_i  \over r^4}~~~,
\quad\quad
b_{ij} = 2\epsilon_{ijkl} {Q_5 x_k \hat f_l \over r^4}~, 
 \label{grabeNS}
\eea
for the NSNS fields and
\bea
C^{(0)} &=& {2 Q_5 \hat f_i x_i\over r^4}  ~,\qquad
C^{(2)}_{ti} =  -\epsilon_{ijkl}{Q_5 x_j \hat f_{kl}\over r^4}\,,\quad C^{(2)}_{yi}=-{2  Q_5 \hat f_{ij}x_j\over r^4}~,\nn\\
C^{(4)}_{tyij}&=& -2\epsilon_{ijkl} {Q_5 x_k \hat f_l \over r^4}\,,\quad C^{(4)}_{abcd}=\epsilon_{abcd} {2 Q_5\hat f_i x_i \over r^4}\,,
\label{grabeR}
\eea
for the RR fields.
 
The simplest example of this type of configurations is provided by a
circular profile in the plane $1,2$
\be
f_1(v)= a \cos {2\pi w v\over L}\,,\quad f_2(v)=a \sin {2\pi w v\over
  L}\,, 
\label{circular}
\ee
while $f_{\hat a}$, $f$ and all remaining
components of $f_i$ are zero. The Eqs.~(\ref{d1charge})
and~(\ref{length}) relate the amplitude $a$ of the profile to $Q_1$ and
$Q_5$
\be
a={\sqrt{Q_1 Q_5}\over w R}\,.
\ee
In this case one can compute exactly all functions in~\eqref{deff},
but for our purposes it is sufficient to look at the order captured
by~(\ref{r31}-\ref{r33}). By using~\eqref{circular} we find
\be
\hat f_{12}=-\hat f_{21}=-{a^2\over 2} {2\pi w\over L}=-{Q_1\over 2 w R}\,,\quad
\hat f_{\hat{a}j} =\hat f_{j}=0\,.
\ee
By using this result in~\eqref{grabeNS} and~\eqref{grabeR} and the
quantization rule~\eqref{quant}, one can see that all terms of order
$1/r^3$ are proportional to $n_1 n_5$.  This clearly suggests that the
microscopic origin of these contributions is related to string
diagrams involving both the D1 and the D5-branes at the same time.

%%%%%%%%%
 
\sect{String vertex operators}

\subsection{Open string vertices}

A simple superposition of D1 and D5-branes does not represent a real
bound state. From the conformal field theory point of view, this is
signalled by the presence of massless string states describing the
relative position of the two stacks of D-branes. We can lift these
modes by giving a non-trivial vacuum expectation value to the open
strings stretched between the two sets of D-branes. The effects of the
condensate can be described in terms of string amplitudes with the
insertions of twisted open string vertices.  The insertion of a
twisted open string state on the boundary of the disk flips a D1
boundary into a D5 and viceversa.  From the supergravity solution the
leading deviation from the naive D1D5 geometry arises at order $n_1
n_5$ and therefore we need two twisted open string insertions.
 
 There are two choices for the open string condensates depending on
whether we excite states from the Neveu-Schwarz (NS) or the Ramond (R)
sector of the open string theory. Turning a vacuum expectation value
for the NS fields generates a non-trivial profile for the open string
photons and can be interpreted as an instanton solution along the ND
directions, see~\cite{Billo:2002hm}. Here our setup is different in
two respects: first the ND directions are compactified on a $T^4$,
second we make a complementary choice for the open string condensate
by turning on the states in the Ramond sector. As we will see, such a
condensate does not generate any a non-trivial gauge (open string)
profile, but only a supergravity (closed string) backreaction,
reproducing the leading asymptotics of the microstate solutions. We
restrict ourselves to systems with only two charges, ({\em i.e.} we
set to zero the quantized momentum along the Neumann direction $S^1$,
that would appear as an additional charge) and so we focus on the open
string states at zero momentum.

We denote the 10D coordinates $(x^{\hat M},\psi^{\hat M})$ with $\hat
M=t,y,1,..8$.  It is convenient to parametrize the coordinates in
terms of the light cone directions $Z^\pm$ and of four complex
variables $Z^I$:
\beq\label{r2c}
Z^\pm \equiv \ft{1}{\sqrt{2}} (y\pm t)~~,\qquad
Z^{n=1,..4}= \ft{1}{\sqrt{2}} (x^{2n-1}+\ii x^{2n}).
\eeq  
We will collectively denote the NN and DD directions by $Z^I$,
$I=+,-,1,\bar 1, 2,\bar 2$ and the mixed ND directions $Z^a$,
$a=3,\bar 3,4 ,\bar 4$.  Indices $I,J,\dots$ and $a,b,\dots$
label the vector representations of the $SO(1,5)$ and $SO(4)$ Lorentz
groups acting on the NN/DD and ND/DN planes respectively.  
In our conventions, the 10D Majorana-Weyl spinors
$\Theta_{\hat A}$ satisfy $\Gamma_{(10)} \Theta_{\hat A}=-\Theta_{\hat
A}$, where $\Gamma_{(10)}=\Gamma^0_{(10)}\Gamma^y_{(10)} \Gamma^1_{(10)}
\ldots \Gamma^8_{(10)}$. These spinors decompose with respect to the
$SO(1,5)\times SO(4)$ as
\be
\Theta_{\hat A}= \{ \Theta_{A}^{~\dot\alpha} ; \Theta^{A \alpha} \} \,,
\ee
where upper and lower indices $A,B,\dots=1,\ldots,4$ denote Weyl
$SO(1,5)$ spinors of opposite chirality; similarly $\alpha,~
\dot{\alpha}=1,2$ are Weyl spinor indices of opposite chirality for
the $SO(4)$ group acting along the ND $T^4$ directions. We decompose
the 10D Gamma matrices as follows
\beq
\Gamma^a_{(10)} = 1_{(6)} \otimes \gamma^a~~,\qquad
\Gamma^I_{(10)} = \Gamma^I \otimes \gamma^{ND}~,
\eeq
where we use simply $\Gamma^I$ for the 6D Gamma matrices and
\bea
(\gamma^{ND})_{\dot\alpha}^{\dot\beta} &=&
(\prod_a \gamma^a)_{\dot\alpha}^{\dot\beta} = 
-\delta_{\dot\alpha}^{\dot\beta}
~,~~~  (\gamma^{ND})_{\alpha}^{\beta}=
(\prod_a \gamma^a)_{\alpha}^{\beta} = 
\delta_{\alpha}^{\beta} ~,\nn\\
(\Gamma)_{A}^{~B} &=&
(\prod_I \Gamma^I)_{A}^{~B}= 
-\delta_{A}^{~B}
~,~~~
(\Gamma)^{A}_{~B}=(\prod_I \Gamma^I)^{A}_{~B}=
\delta^{A}_{~B}\,.
\label{gGamma}
\eea
Instead of the 6D Gamma matrices, we will often use the chiral
components such as $ (C \Gamma^{I_1..I_{2n-1}})_{AB}$, where $C$ is the
6D charge conjugation matrix\footnote{$C$ is related to the 10D and 4D
charge conjugation matrices by $C_{10}=C\otimes C_4$.} satisfying ${}^{\rm
t}\Gamma^I= -C \Gamma^I C^{-1}$.

We will consider string amplitudes describing a closed string emission
from a mixed D1D5 disk with {\it twisted} open string vertex
insertions.  We refer the reader to \cite{Billo:2008} for a general
discussion of closed string amplitudes on disks with mixed boundary
conditions and for details on the conventions we follow here.  Twisted
open string vertices are associated to string states stretched between
D-branes with different boundary conditions and involve bosonic twist
fields. Since there is no simple conformal field theory description
when these fields have a non-zero expectation value, we will work
perturbatively, and consider the leading contribution coming from the
insertion of a single pair of twisted open string vertices. The open
string vertices flip the boundary conditions on the disk from D1 to D5
types and viceversa and therefore split the disk boundary into two
portions with D1 and D5 boundary conditions respectively.  We restrict
ourselves to massless physical states describing the lowest
excitations of open strings stretched between D1 and D5 branes (for
more details see, for example, \cite{Billo:2002hm}).  In the NS
sector, these states are generated by the zero-modes of the $T^4$
fermions, $\psi^a$, and hence form a spinor representation of
$SO(4)$. The associated vertex operators are
\beq\label{Vop14} 
V_{w} = w_{\dot\alpha}\ex{-{\varphi}} S^{\dot \alpha} \,
 \Delta\,, \quad\qquad ~~~~~~~~
V_{\bar w} =\bar w_{\dot \alpha} \ex{-{\varphi}}  
S^{\dot\alpha} \, \dbar\,,
\eeq
with $w_{\dot\alpha}, \bar{w}_{\dot\alpha}$ denoting $n_1\times n_5$
and $n_5\times n_1$ Chan-Paton matrices respectively.  In the R
sector, it is the fermions along the $I$ directions that have zero
modes, and hence the physical states form a spinor representation of
$SO(1,5)$. They correspond to the vertex operators
\beq\label{Vop15} 
V_{\mu} = \mu^{A} \ex{-{\varphi\over 2}} S_A \,
 \Delta \,, \quad\qquad ~~~~~~~~
V_{\bar \mu} =\bar \mu^{A} \ex{-{\varphi\over 2}}  S_A \, \dbar \,,
\eeq
where the Chan-Paton matrices $\mu^A$ and $\bar{\mu}^A$ have $n_1\times n_5$ and $n_5\times n_1$ components respectively. 
In (\ref{Vop14},\ref{Vop15}) and below, we denote by $\varphi$ the
free boson appearing in the bosonized language of the worldsheet
superghost $(\beta,\gamma)$.  $\Delta$ is the bosonic twist operator
with conformal dimension ${1\over 4}$, that acts along the four mixed
ND directions and changes the boundary conditions from Neumann to
Dirichlet and viceversa.  $S^\alpha$ and $S_A$ are the $SO(4)$ and
$SO(1,5)$ spin fields. After bosonization these spin fields are simply
exponentials of free bosons.  If one introduces the bosons $h_{\bf I}=
( h_+, h_1, h_2)$ and $h_{\bf a} = (h_3, h_4)$ associated to the
$SO(1,5)$ and $SO(4)$ fermions respectively, fermions and spin fields
are given by\footnote{In our conventions $h_{\bf I} (z_1) h_{\bf J}
(z_2) \sim -\delta_{\bf IJ} \log z_{12}$, $h_{\bf a} (z_1) h_{\bf b}
(z_2) \sim -\delta_{\bf ab} \log z_{12}$.  $h_{\bf I \neq +}$ are real
fields and $h_+$ purely imaginary \cite{Polchinski:1998rr}. }
\begin{align}
\label{spinR}
 & S^A = S^{\vec \epsilon^A} 
=\ex{\frac{\ii}{2} \epsilon^{A \bf I} h_{\bf I} } \,,
%\quad\quad  ~~~~~~
& S_A = S^{\vec \epsilon_A}   
=\ex{\frac{\ii}{2} \epsilon^{\bf I}_A h_{\bf I} } \, , 
%\quazd\quad  
& &\psi^{\bf I} = \ex{\ii h_{\bf I} } \,,
%\quad\quad  
&& \bar\psi^{\bf I} = \ex{-\ii h_{\bf I} }\,, \\
% \label{spinNS}
& S^{\dot{\alpha}} = S^{\vec \epsilon_{\dot \alpha} } 
=\ex{\frac{\ii}{2} \epsilon^{\bf a}_{\dot{\alpha}}  h_{\bf a}}\,,
%\quad\quad  ~~~~~~
& S^{{\alpha}} = S^{\vec \epsilon_{ \alpha} } 
=\ex{\frac{\ii}{2}  \epsilon^{\bf a}_{\alpha} h_{\bf a} }\,,
%\quad\quad~~  
& & \psi^{\bf a} = \ex{\ii h_{\bf a} } \,,
%\quad\quad~~  
& &\bar \psi^{\bf a} = \ex{-\ii h_{\bf a} }\,, \nn
\end{align}
with ${\bf I}=+,1,2$, ${\bf a}=1,2$ running over the holomorphic
components and $A,\alpha,\dot{\alpha}$ running over the spinor
components corresponding to the following choices of signs
\bea
\vec \epsilon_A &=&  \{ (---), (-++),(+-+),(++-) \} \,,\nn\\
\vec \epsilon^A &=&  \{ (+++), (+--),(-+-),(--+) \} \,,\nn\\
\vec \epsilon_\alpha &=&  \{ (++), (--) \} \,,\nn\\
\vec \epsilon_{\dot \alpha} &=&  \{ (+-), (-+) \}\,. 
\eea
 As we mentioned before, we focus on open string condensates involving
 only states from the Ramond sector.  Notice that states in the Ramond
 sector will break the $SO(4)$ symmetry of the DD directions
 $\mathbb{R}^4$, while they are invariant under the $SO(4)$ acting on
 the compact $T^4$ torus. They will be then associated with those
 supergravity solutions in \eqref{generald1d5} with ${\cal A}_{\hat
 a}=0$.  The most general condensate of Ramond open strings can be
 written as:
\be\label{mubarmu}
\bar \mu^A \, \mu^B=v_I (C\Gamma^I)^{[AB]}+
\frac{1}{3!} \,v_{IJK} (C\Gamma^{IJK})^{(AB)}~,
\ee
where the parenthesis on the indices $A, B$ are meant to remind that
the first term is automatically antisymmetric, while the second one is
symmetric. Thus the open string bispinor condensate is specified by a
one-form $v_I$ and an self-dual three-form $v_{IJK}$. The self-duality
of $v_{IJK}$ follows from $\bar\mu^A$ and $\mu^B$ having definite 6D
chirality and can be written as
\beq\label{sdv}
v_{IJK}= \frac{1}{3!} \epsilon_{IJKLMN} v^{LMN}~.
\eeq
We will focus on symmetric open string condensates satisfying $v_I=0$.
In fact, turning on a vacuum expectation value for $v_I$ would
generate a tadpole for the D1D1 and D5D5 untwisted
fields\footnote{This can be seen by computing the three point function
$\langle V_\mu V_A V_{\bar \mu}\rangle$ with $V_A=e^{-\varphi} \psi^I
$.} $A^{(1)}_I$, $A^{(5)}_I$. This is not the case of $v_{IJK}$
components that do not have any trilinear coupling with open string
states and therefore one can turn on a condensate $\langle v_{IJK}
\rangle \neq 0$ without generating a tadpole in the open string
theory.  We will consider D1D5 geometries generated by the mixed disks
involving these non-trivial condensates , i.e.
\be
\bar\mu^{(A}\mu^{B)}= \frac{1}{3!} \, v_{IJK} (C\Gamma^{IJK})^{AB}  \,,
 \qquad 
v_{IJK} = -\ft18  \, \bar \mu^A (\Gamma_{IJK}C^{-1})_{AB} \mu^B  \,.
\label{conds2}
\ee
It has to be remembered that the spinors $\bar \mu^A$ and $\mu^B$ carry $n_5\times n_1$ and $n_1\times n_5$ Chan-Paton indices, and hence the condensate $\bar\mu^{A}\mu^{B}$ has to be thought of as the vev for the sum
\be
  \sum_{m=1}^{n_1} \sum_{n=1}^{n_5}  \bar\mu^{A}_{m n}\,  \mu^{B}_{n m}\,,
\ee
which, for generic choices of the Chan-Paton factors, is of order $n_1
n_5$. We thus see that the amplitudes we compute are of the same order
in $n_1 n_5$ as the gravity terms \eqref{grabeNS} and \eqref{grabeR}.
In the following we will assume that the second identity in
~\eqref{conds2} already contains the trace over Chan-Paton indices,
and thus that $v_{IJK}$ is of order $n_1 n_5$.

\subsection{Closed string vertex operators}

The vertex operators in the closed string sector
are\footnote{For our purposes we will not need the absolute
normalization of the vertex operators; the relative normalization is
determined by requiring that the NSNS and RR states are related to the
canonically normalized fields with the same proportionality constant
(for details see \cite{Billo:1998vr}).}
\begin{eqnarray}
W_{NS}& = & {\cal G}_{\hat{M}\hat{N}}
 \left(\partial X^{\hat{M}}_L -
\ii  k_L \cdot\psi \psi^{\hat{M}} \right)
 \,\ex{ \ii  k_L X_L} (z)
\, \tilde\psi^{\hat{N}} 
\ex{-\tilde\varphi} \ex{ \ii  k_R X_R }(\bar z)\,,
\label{clNS0} \\ 
\label{clR0}
W_{R} &=& \frac{1}{4\sqrt{2}}\,  {\cal F}_{\hat{A}\hat{B}} 
 \ex{-{\varphi\over 2}}\, S^{\hat{A}} \, \ex{ \ii  k_L X_L} (z)
 \,\ex{-\tilde {\varphi\over 2}} \, \tilde S^{\hat{B}}\, 
 \ex{ \ii  k_R X_R}(\bar z) ~,
\end{eqnarray}
where all hatted indices are ten dimensional and ${\cal
G}_{\hat{M}\hat{N}}$ contains the 10D metric, the NSNS 2-form, and
the dilaton, while ${\cal F}_{\hat{A}\hat{B}}$ is a RR field strength which
can be expanded on a basis of ten dimensional Gamma matrices and
contains a 1, a 3 and a self-dual 5-form
\be
{\cal F}_{\hat A \hat B} =\sum_{n=1,3,5}  
  \frac{1}{n!}
F^{(n)}_{\hat M_1..\hat M_n}(C_{10}\Gamma_{(10)}^{\hat M_1..\hat M_n})_{\hat{A}\hat{B}} \,.  \label{f135} 
\ee
Closed string vertices are
separately normal ordered in the left and right moving terms. This is
important in disk amplitudes where the left and right moving fields
are identified
\be \tilde \varphi=\varphi~~~,~~~~ 
X_R^{\hat{M}} =R^{\hat{M}}_{\;\hat{N}} \, 
X_L^{\hat{N}}~~~,~~~~ \tilde \psi^{\hat{M}} =
R^{\hat{M}}_{\;\hat{N}} \, \psi^{\hat{N}} ~~~,~~~~
\tilde S^{\hat{A}} = R^{\hat A}_{\;\hat B} \, S^{\hat B} 
%\equiv (\prod_{i=1}^4 \Gamma^i)^A_B \, S^B 
%~,~~ \tilde S^{\alpha} = R^{\alpha}_{\;\beta}S^{\beta}
~.
\label{condm}
\ee
The identification matrix $R$ depends on the D-brane boundary
conditions: $R^{\hat{M}}_{\;\hat{N}}$ is a diagonal matrix with $-1$
along the Dirichlet directions and $+1$ otherwise, while the
identification matrices with spinor indices is, up to a sign, the
product of the (ten dimensional) Gamma matrices along the Neumann
directions. In our case, we have two possible choices for $R$,
depending whether we use the boundary conditions of the D1 or the D5
branes
\beq\label{RRR}
R_{\mathrm{D1}}= \Gamma^{ty}_{(10)}= \Gamma^{+-}_{(10)}~,\qquad
 R_{\mathrm{D5}} = \Gamma^{ty5678}_{(10)} =
-\Gamma^{+- 3 \bar 3 4 \bar 4 }_{(10)} \,,
\eeq 
with $\Gamma^{5678}_{(10)}$ the chirality operator along $T^4$ written in terms
of the 10D Gamma matrices.  The final result of each correlator should
not depend on the particular choice made for $R$. Here we will always
take $R=R_{D_1}$.

As we are interested in configurations that are translationally
invariant both along the world sheet directions $t$ and $y$ and the
$T^4$ directions, we take the closed string momentum vector along
$\mathbb{R}^4$: $k^i_L=k^i_R=k^i/2$ . The asymptotics of the D1D5
geometries generated by the mixed disks will be captured by the
leading term in the expansion of the string amplitude for small $k_i$.
We are interested only on the leading term in the momentum expansion,
and therefore we can set to zero the momentum in the exponential of
closed string vertices.  Finally, since the open string condensate
$\bar \mu^{(A} \mu^{B)}$ under consideration is invariant under the
$SO(4)$ Lorentz group of the $T^4$ torus, we can restrict ourselves to
$SO(4)$ invariant components ${\cal G}_{IJ}$, ${\cal
F}_{AB[\dot\alpha \dot\beta]}$ , ${\cal
F}^{AB[\alpha \beta]}$ in~(\ref{clNS0}) and~(\ref{clR0}). In
addition the RR components ${\cal F}^{AB[\alpha \beta]}$ can be
discarded by noticing that the only $SO(6)$ singlet $\epsilon_{ABCD}
\mu^{A} \bar\mu^{B}\, {\cal F}^{CD [\alpha \beta]} $ vanishes for
the symmetric open string condensate $\bar \mu^{(A} \mu^{B)}$
considered here. 

\sect{Microstate geometries from disk amplitudes}

In this section we show how the small $g_s$, long distance behavior of
the microstate solutions~\eqref{generald1d5} is reproduced by using
string amplitudes and the microscopic description of D-branes.
In particular, we compute the emission of one closed string state from
various mixed disks ({\em i.e.} disks that have half of their boundary
along the D1-branes and the other half along the D5-branes) and
extract from these amplitudes the leading deviation of the
geometries~\eqref{generald1d5} from the naive D1D5 metric, which
appears at order $1/r^3$ in the large $r$ limit. We will restrict
ourselves to open string condensates describing D1D5 geometries with
trivial profile along the compact $T^4$ directions, i.e.  ${\cal
A}_{\hat a}={\cal B}_{\hat a ij}=0$. The
correlators we compute reproduce the terms in the geometries~\eqref{generald1d5}
of order $1/r^3$, which are captured by the functions $A_i$, ${\cal
A}$, and their duals $B_i$, ${\cal B}_{ij}$ given in
eqs. (\ref{r31})-(\ref{r34}).

The emission of closed string states from a D1D5 system with an open
string condensate turned on can be described by computing string
diagrams with a closed string state and some number of open string
insertions. The leading contribution comes from the disk with one
closed and two twisted open string vertices. The insertion of two
twisted open string vertices divides the disk into two boundaries with
D1 and D5 boundary conditions. The relevant correlator can be written
as
\begin{equation}\label{c}
{\cal A}_W = \int \frac{\prod_{i=1}^4dz_i}{dV_{\mathrm{CKG}}} \,
\langle V_\mu%_\mu
(z_1) 
W_{NS,R}(z_2, z_3) 
V_{\bar \mu} (z_4) \rangle~,
\end{equation} 
where $W_{NS,R}$ represents the emitted closed string. The closed
string operators to be inserted in~\eqref{c} need to have total
superghost charge $-1$ so that it compensates, together with the
superghost charge of the open string vertices, the $-2$ background
charge of the disk.  The relevant vertex operators are given by
(\ref{Vop15},\ref{clNS0},\ref{clR0}). The open string variables, denoted
by $z_1$ and $z_4$, are integrated on the real axis while the closed
string variables $z_2=z$ and $z_3=\bar z$ are the complex conjugate of
each other and $z$ must be integrated over the upper half complex plane.  The
$\mathrm{SL}(2;\mathbb{R})$ projective invariance is fixed explicitly
by choosing
\begin{equation}
\frac{\prod_{i=1}^4dz_i}{dV_{\mathrm{CKG}}}= d\omega \,
\big(z_{13}z_{24}\big)^2
\label{dz}
\end{equation}
where $\omega$ is the ratio\footnote{The integral measure follows from
$\frac{\prod_{i=1}^4dz_i}{dV_{\mathrm{CKG}}}=dz_2 \langle c(z_1)
c(z_3) c(z_4)\rangle=dz_2 z_{13} z_{34} z_{41} =d\omega \,
\big(z_{13}z_{24}\big)^2$. The other two ratios that can be built from
$\omega$ are $1-\omega=\frac{z_{23}z_{14}}{z_{13}z_{24}}$ and
${1-\omega\over \omega}=\frac{z_{23}z_{14}}{z_{12}z_{34}}$.  }
\begin{equation}
\omega= \frac{z_{12}z_{34}}{z_{13}z_{24}}~, 
\label{omega}
\end{equation}
with $z_{ij}=z_i-z_j$. Notice that $\omega$ is a pure phase since
$z_2$ and $z_3$ are complex conjugate of each other, while $z_1$
and $z_4$ are real. 

\subsection{NSNS amplitude  }

Let us start from the emission of the NSNS state  
\begin{equation}\label{NS00}
{\cal A}_{NS} = 
\int \frac{\prod_{i=1}^4dz_i}{dV_{\mathrm{CKG}}} \,
\langle V_\mu(z_1) W_{NS}(z_2, z_3) V_{\bar\mu}(z_4) \rangle~.
\end{equation}
The transversality condition $k^{I} {\cal G}_{IJ}=0$ implies that the
term of the closed vertex~\eqref{clNS0} proportional to $\partial
X^{\hat{M}}$ does not contribute to the correlator: only the term
proportional to $k_L \cdot\psi \psi^{\hat{M}}$ contributes. 
Only one Lorentz invariant can be built out of $k_K$, $({\cal G}
R)_{IJ}$ and $v_{IJK}$ and so the form of the string amplitude is
determined from Lorentz invariance up to a constant ${\cal I}$
\be \label{ampf0}
{\cal A}_{\rm NS}=- k_K\, ({\cal G} R)_{IJ}\,  \bar \mu^A
\mu^B\,(\Gamma^{IJK}C^{-1})_{AB} 
\,{\cal I} =8 k_K\, ({\cal G} R)_{IJ}\, v^{IJK}
\,{\cal I}  ~,
\ee
where the last equality follows from the second relation
in~(\ref{conds2}).  It is important to stress that the CFT correlator
is $SO(1,5)$ invariant, and only the explicit form of the matrix $R$
breaks this invariance down to $SO(1,1)\times SO(4)$.  The constant
${\cal I}$ has to be computed from the explicit evaluation of the CFT
correlator. To evaluate ${\cal I}$, we specify to the term in
(\ref{ampf0}) with $K=1,I= 2,J=+$ and $\vec{\epsilon}_A=\vec{\epsilon}_B= (---)$.
Notice that for this choice of open string polarizations, the indices
$A,B$ are automatically symmetric consistently with the desired form
of the $\bar \mu \mu$-condensate (see Eq.~\eqref{conds2}). Charge
conservation implies that only the cubic fermionic term
of the vertex~\eqref{clNS0}, in which all the three fermionic fields
carry $SO(1,5)$ indices, can contribute to this amplitude and
therefore one can write
\be
W_{NS}=-\ii \,\frac{k_K}{2}\, ({\cal G}R)_{IJ} \, 
\psi^K \psi^I (z_2) \psi^J (z_3)\,.
\ee
Thus the relevant correlators are
\bea 
\Big\langle \Delta (z_1)\, \dbar (z_4) \Big\rangle &=& 
z_{14}^{-{1\over 2}}~, \nn\\
\langle  \ex{-{\varphi\over 2}} (z_1)  \, \ex{-\varphi} (z_3)  \, 
\ex{-{\varphi\over 2}} (z_4)  \, \rangle 
 &=&  (z_{13} z_{34})^{-{1\over 2}} \, z_{14}^{-{1\over 4}}~, \nn\\
   \langle  S^{---} (z_1)  
\,  \psi^1 \psi^2 (z_2)  \psi^+(z_3) 
\,  S^{---} (z_4)
\rangle  &=&    (z_{12} z_{24}  )^{-1}    (z_{13} z_{34}) ^{-{1\over
    2}}        z_{14} ^{ {3\over 4}} ~.
 \label{corr4}
\eea 
The last equation can be written in the covariant form\footnote{The
origin of the factors of $\sqrt{2}$ lies in the OPE's of the
$\psi^I$'s and the twist fields~\cite{Kostelecky:1986xg,Billo:2002hm}:
\be
\psi^I(z) S_A(0) \sim {1\over \sqrt{2}} {\Gamma^I_{AB} S^B(0)\over z^{1/2}}\nonumber\,. 
\ee
Note that that equation~\eqref{corr4p} is verified also for other
choices of the polarizations $\vec{\epsilon}_{A,B}$. Take, for
example, $\vec{\epsilon}_A= (---)$, $\vec{\epsilon}_B = (-++)$ and
$I=+$, $J=1$, $K=\bar 1$. Then on one side one has
\be
  \langle  S^{(---} (z_1)  
 \,  \psi^+\psi^1 (z_2)  \psi^{\bar 1}(z_3) 
\,  S^{-++)} (z_4) \rangle  =   \frac{1}{2}
(z_{12} z_{24}  )^{-1} (z_{13}
  z_{34})^{-{1\over 2}}  z_{14} ^{ {3\over 4}}  \,,\nonumber
\ee
and on the other side
\be
(\Gamma^{+ 1 \bar 1})_{---,-++} = {1\over 2} (\Gamma^+ \Gamma^1 \Gamma^{\bar 1})_{---,-++}=
\sqrt{2}\,,\nonumber
\ee
which agrees with~\eqref{corr4p}.
 }
\be
  \langle  S_{(A} (z_1)  
 \,  \psi^I\psi^J (z_2)  \psi^K(z_3) 
\,  S_{B)} (z_4) \rangle  =  \frac{1}{2\sqrt{2}}
(\Gamma^{IJK}C^{-1})_{AB} \, (z_{12} z_{24}  )^{-1} (z_{13}
  z_{34})^{-{1\over 2}}  z_{14} ^{ {3\over 4}}  \,.
\label{corr4p} 
 \ee
By inserting Eqs.~(\ref{dz},\ref{corr4},\ref{corr4p}) into \ref{NS00},
one finds
\be
{\cal I}=\frac{\ii}{4\sqrt{2}} \int {d\omega\over \omega }=
-\frac{\pi}{2\sqrt{2}} 
\ee
with the integral running over the unitary circle.  Substituting the
value of ${\cal I}$ in (\ref{ampf0}), one finds that the string
amplitude is
\be  
{\cal A}_{\rm NS} =- 2\sqrt{2} \pi   \, k_K\, ({\cal G} R)_{IJ}\,
v^{IJK}   \,.
\label{ampf0I} 
\ee
As we will now show, this amplitude exactly reproduces the $1/r^3$
contributions to the metric and the B-field in~\eqref{generald1d5}
associated with functions $A_i$, $B_i$, ${\cal A}$ and $\cal B$. In
order to identify the fields appearing in~\eqref{ampf0I}, we need to
decompose the $SO(1,5)$ vector indices into $SO(1,1) \times SO(4)$
indices ($I=(t,y,i)$, where, as before, $i=1,..4$ label the
$\mathbb{R}^4$ Dirichlet directions and $t,y$ the Neumann directions).
Notice that only antisymmetric components $({\cal G} R)_{[JK]}$
contribute to the amplitude.  Using the fact that the matrix $R$ is
$+1$ along $t,y$ and $-1$ otherwise, one finds that the matrix $({\cal
G} R)_{JK}$ is antisymmetric if and only if ${\cal G}$ is
antisymmetric (B-field) and the $(J,K)$ indices are of the same type
(i.e. $(t,y)$ or $(i,j)$), or ${\cal G}$ is symmetric (metric) and the
$(J,K)$ indices are of the different type (i.e. $(t,i)$ or $(y,i)$).
This implies that only the components $g_{t i}, g_{y i}$ and
$b_{ty},b_{kl}$ are emitted from the mixed disk.  In addition we
recall that the momentum $k_I$ of the closed string is non-zero only
in the Dirichlet directions $i$.  The amplitude (\ref{ampf0I}) can
then be written as
\bea  
{\cal A}_{\rm NS} &=& 4\sqrt{2} \pi  \left(  k_j \, g_{ti}v^{ t i j} \,
+ k_j g_{y i} v^{yij} -  k_i \, b_{ty}v^{t y i }    \,   
+ \ft12 k_k b_{ ij} v^{ijk}\right)\nn\\ 
&=&4\sqrt{2} \pi  \left(  k^j \, g^{ti}v_{ t i j} \, +\ft12 k^j g^{y i}
\epsilon_{  i j kl } v_{tkl}  -
k^i \, b^{ty}v_{  t y i } \, -\ft12  k^k b^{ ij} \epsilon_{  ij kl
} v_{tyl}\right)  \,,
\label{ampf0II} 
\eea
where in the second line we used the self-duality conditions of the
3-form $v^{IJK}$ given by
\be \label{vtij}
v_{yij} =  \frac 12 \epsilon_{ijkl} v_{tkl}~ \qquad v_{ijk} = -
\epsilon_{ijkl} v_{tyl} \,.
\ee
From~\eqref{ampf0II} we can read the profile of the induced metric and
$B$ field, for instance 
\be\label{vans}
g_{t i}(k) =\frac 12 {\delta {\cal A}_{\rm NS} \over\delta g^{t i}}  
=  2\sqrt{2} \pi \, k_j v_{tij} ~,\qquad
b_{t y}(k) ={\delta {\cal A}_{\rm NS} \over\delta b^{t y}} 
=  -4 \sqrt{2} \pi \, k_j v_{tyj} ~.
\ee
As in~\cite{DiVecchia:1997pr}, the space-time configuration associated
with a closed string emission amplitude is obtained by multiplying the
derivative of the amplitude with respect to the closed string field by
a free propagator and taking the Fourier transform. In general for a
field $a_{\mu_1\ldots\mu_n}$ we have
\beq\label{fourier}
a_{\mu_1\ldots\mu_n}(x) = \int \frac{d^4 k}{(2\pi)^4}
\left(-\frac{\ii}{k^2}\right)  
a_{\mu_1\ldots\mu_n}(k)\,
\ex{\ii k x} ~,
\eeq
with $a_{\mu_1\ldots\mu_n}(k)$ given in terms of derivatives of ${\cal
A}$ as in~(\ref{vans}). In our case, the Fourier transform has the
following form
\be\label{fti}
\int \frac{d^4 k}{(2\pi)^4} \left(- \frac{\ii }{k^2} \right) 
k_j \ex{\ii k x} = - \frac{\partial}{\partial x^j}
\int  \frac{d^4 k}{(2\pi)^4}  \frac{1}{k^2}  
\,   \ex{\ii k  x}  =  - { 1\over 2\pi^2}{x_j \over r^4 }~,
\ee
where we used
\begin{equation}
\int \frac{d^4 k}{(2\pi)^4}  \frac{e^{\ii k  x}}{k^2}   =
-\frac{1}{4 \pi^2} \frac{1}{r^2}~.
\end{equation}
 Thus from~\eqref{vans} we get the
following results for the large distance behavior of $g_{ti}$ and
$b_{ty}$ at order $1/r^3$
\beq\label{gti}
g_{ti}(x) = -{\sqrt{2}\over \pi}{x^j v_{tij}\over r^4 } ~~~,\qquad
b_{ty}(x)  = {2\sqrt{2} \over \pi} {x^i v_{tyi}\over r^4}~.
\eeq
We can follow the same steps for the variation of ${\cal A}_{\rm NS}$
in~(\ref{ampf0II}) with the respect to $g_{yi}$ and $b_{ij}$ leading to
\beq \label{gyi}
g_{yi}(x) = -{1\over \sqrt{2}\pi} \epsilon_{ijkl} {x^j v_{tkl}\over r^4 }
~~~,
\quad\quad
b_{ij} (x)=  {2 \sqrt{2}\over \pi}   \epsilon_{ijkl}
 {x^k v_{tyl}\over r^4}~.
\eeq
Comparing the values of $g_{ti}$ and $b_{ty}$ derived above with the
$1/r^3$ terms of the gravity solution \eqref{grabeNS}, one fixes the
identification between the string condensate parameter $v_{IJK}$ and
the parameters ${\hat f}_{ij}$ and ${\hat f}_i$ that characterize the
gravity solution at this order:%
\be
Q_5 \hat f_{ij} = \frac{1}{\sqrt{2}\pi} v_{tij}~ ,\qquad ~~~~~~~~~~ Q_5
\hat f_{i} = \frac{\sqrt{2}}{\pi} v_{tyi} \,.
\label{idenfij} 
\ee 
Using this identification, one verifies that the components $g_{yi}$
and $b_{ij}$ predicted by the string computation agree with the
gravity result \eqref{grabeNS}.

\subsection{RR amplitude}

Let us now consider the amplitude describing the emission of RR
states
\begin{equation}\label{R}
{\cal A}_{R} = 
\int \frac{\prod_{i=1}^4dz_i}{dV_{\mathrm{CKG}}} \,
\langle V_\mu(z_1) W_{R}(z_2, z_3)  V_{\bar\mu}(z_4) \rangle~,
\end{equation} 
with $W_{R}$ given by (\ref{clR0}). As discussed at the end of
Section~3, the open string condensate under analysis contributes only
to the emission of
\beq
\label{clR}
W_{R}^{\rm ef} = \frac{1}{4\sqrt{2}}\,({\cal F}R)_{A B} \epsilon_{\dot\alpha\dot\beta}
 \ex{-{\varphi\over 2}}\, S^{A} \, S^{\dot\alpha} \,  (z)
 \,\ex{- {\varphi\over 2}} \, S^{B}\, S^{\dot \beta}(\bar z) ~,
\eeq
with
\bea
{\cal F}_{AB} &=&   \frac{1}{2\,  4!}  F^{(5)}_{Iabcd}( C_{10}
\Gamma^{Iabcd}_{(10)})_{AB}{}^{\dot\alpha}_{\dot\alpha}+ 
\sum_{n=1,3,5} \frac{1}{2\,  n!}  
F^{(n)}_{I_1..I_n} (C_{10} \,
\Gamma^{I_1..I_n}_{(10)})_{AB}{}^{\dot\alpha}_{\dot\alpha}\nn\\ 
&=&  F^{(5)}_{I 5678}
( C \,\Gamma^{I})_{AB} +\sum_{n=1,3,5} \frac{1}{  n!}   
F^{(n)}_{I_1..I_n} (C \,\Gamma^{I_1..I_n})_{AB} \,, \label{rrforms}
\eea
where, in the last line, we have used $\Gamma^{5678}_{(10)} =
-\Gamma^{3\bar 3 4 \bar 4}_{(10)}=1_{(6)}\otimes (-\gamma^{ND})$ and
that $-\gamma^{ND}$ is $1$ on the indices $\dot{\alpha}$ (see \eqref{gGamma}).
Lorentz invariance again fixes the form of ${\cal A}_R$ up to a
constant
\be
{\cal A}_R = 2 {\cal I} \,\bar \mu^A(C^{-1} {\cal F} R C^{-1})_{AB}\,
\mu^B~, 
\ee
where the factor of two in evidence comes from the trace over the 4D
spinor indices. Notice that both the D1 and the D5 reflections
matrices~\eqref{RRR} reduce in 6D to the expression $R=\Gamma^{ty}$,
since they differs by $\Gamma^{5678}_{(10)}$ which is just the
identity on the spinor components ($\dot\alpha$) entering in this
computation. Due to the symmetry properties of the open string
condensate, this result is non-vanishing only when the Gamma matrices
in ${\cal F}$ and $R$ reconstruct a 3-form~\eqref{conds2}. This is
possible for all terms in~ (\ref{rrforms}), thus 1, 3 and 5-forms
components of the RR field will all contribute to the string
amplitude. We use ${\cal A}_{\rm R}[n]$ to indicate the contribution
of the form of degree $n$. By using again the second relation
in~(\ref{conds2}), we get
\bea
{\cal A}_R[1] &=& 16  \,{\cal I}  \,   
v_{tyi}\,  F^{(1)}_{i} \,,\nn\\
{\cal A}_R[3] &=& 16 \, {\cal I}  \, \left( \ft12 
v_{yij}\, F^{(3)}_{tij}- \ft12 v_{tij}\, F^{(3)}_{yij} \right)
\label{amm135}\,, \\  
{\cal A}_R[5]  &=&  -16  \,{\cal I} \,\left( \ft{1}{3!}
v_{ijk}\, F^{(5)}_{tyijk} - v_{tyi} \, 
F^{(5)}_{i5678}\right) \,.  \nn 
\eea
To evaluate ${\cal I}$, one can take a specific choice of the open and
closed string polarizations; for instance, it is convenient to choose
for $\bar \mu^{A}$, $\mu^{B}$ the following weights $\vec
\epsilon_A=\vec \epsilon_B=(---)$.
The relevant correlators are
\bea
 && \Big\langle
  \Delta (z_1)\, \dbar (z_4) \Big\rangle
  =  z_{14}^{ -{1\over 2}}~, \qquad 
  \langle  S^{\dot\alpha}(z_2)  \,  S^{\dot\beta} (z_3) \rangle =
 z_{23}^{-{1\over 2}} \, \epsilon^{\dot\alpha \dot\beta}
 ~,\qquad
  \langle  \prod_i e^{-{\varphi\over 2}} (z_i)    \, \rangle 
= \prod_{i<j} z_{ij}^{-{1\over 4}}~,  \nn\\
 &&\langle  S^{---} (z_1)  \, S^{+++} (z_2)  S^{+++}(z_3) \,  S^{---}
 (z_4)\rangle  = \left({z_{14} z_{23}   \over  z_{12} z_{13}
 z_{24}z_{34} }\right )^{{3\over 4}}
~. 
\eea 
Assembling the various correlators together, and using again the
measure~\eqref{dz}, one finds
\be
{\cal I} =\frac{1}{4\sqrt{2}}\, \int {d\omega\over \omega}=\frac{\pi
  \ii}{2\sqrt{2}}\,. 
\ee
We can now show that the amplitude~\eqref{amm135} contains the $1/r^3$
contributions to the RR fields in the solution~\eqref{generald1d5}
that are characterized by the functions $A_i$, $B_i$, ${\cal A}$ and
${\cal B}$. We first need to rewrite the result~\eqref{amm135} in
terms of the gauge potentials $C^{(n-1)}$ by using
\be
F^{(n)}_{I_1..I_n}=n \,\ii\, k_{[I_1} \,C^{(n-1)}_{I_2..I_n]}~.
\ee
So we get
\bea\label{amm136}
{\cal A}_{\rm R}[1]  &=& -4\sqrt{2} \pi k_i C^{(0)}    v_{tyi}\,,\nn\\
{\cal A}_{\rm R}[3]  &=& -4\sqrt{2}\pi  (\ft12 k_j C^{(2)}_{ti} v_{y i
  j}-\ft12  k_j C^{(2)}_{yi} v_{ t i  j} ) \nn\\ &=&  
 + 4\sqrt{2} \pi (\ft14 k_j C^{(2) ti} \epsilon_{ijkl} v_{tkl} + \ft12
 k_j C^{(2) yi} v_{ t i  j} )\,, \nn\\ 
{\cal A}_{\rm R}[5]  &=& 4\sqrt{2}\pi (\ft{1}{2}\,k^i  C^{(4)}_{tyjk} v_{i
  j k }- \, k^i C^{(4)}_{5678}\,v_{tyi})\nn\\ 
&=& 4\sqrt{2}\pi (\ft{1}{\,2}\,  \epsilon_{ijkl} k^i C^{(4) tyjk} 
v_{tyl} - \, k^i C^{(4) 5678}\, v_{tyi})~.
\eea
We then extract from the amplitude the gauge field profile
\be
C^{(n)}_{\mu_1\ldots\mu_n}(k)={\delta {\cal A}_{\rm R}\over \delta C^{(n) \mu_1\ldots\mu_n}}\quad (\mu_1<\mu_2\ldots<\mu_n)\,,
\ee
attach a free propagator to each profile and take the Fourier
transform, as explained in~\eqref{fourier}. Hence~\eqref{amm136}
yields the following results for the large distance behavior of the RR
fields
\bea
C^{(0)}(x)&=& {2\sqrt{2}\over \pi} {x_i v_{tyi}\over r^4}\,,\nn\\
C^{(2)}_{ti}(x)&=& -{1\over \sqrt{2}\pi}\epsilon_{ijkl}{x_j v_{tkl}\over
  r^4}\,,\quad C^{(2)}_{yi}(x)= - {\sqrt{2}\over \pi}{x_j v_{tij}\over r^4}\,,\nn\\ 
C^{(4)}_{tyij}(x) &=& -{2\sqrt{2} \over \pi}\epsilon_{ijkl} {x_k v_{tyl}\over
  r^4}\,,\quad C^{(4)}_{5678}(x)= {2\sqrt{2}\over \pi} {x_i
  v_{tyi}\over r^4} \,.
\eea
Using the identifications~\eqref{idenfij}, the values above exactly
reproduce the supergravity result ~\eqref{grabeR}.

\sect{Conclusions}

We have shown how the asymptotic expansion of the 2-charge fuzzball
geometries (\ref{generald1d5}) is reproduced by computing string
amplitudes for the emission of a closed string state from a disk with
mixed D1D5 boundary conditions. Each fuzzball geometry is completely
determined by a curve $f_A(v)$ that captures how different the
solution is from the naive D1D5 superposition. Microscopically the
information about the curve $f_A(v)$ is encoded in a condensate for
the open strings stretched between the two types of D-branes. In order
to derive the exact dictionary between $f_A(v)$ and the string
condensate one should compute the closed string emission from a disk
in presence of a finite value for the open condensate. This is a
challenging task since the open string states stretched between the D1
and the D5 branes contain twist fields. However we could explicitly
check this dictionary by treating the open condensates perturbatively:
in the large distance limit the $1/r^3$ terms of the fuzzball
solutions match the gravitational
backreaction of the D-brane system when the open string condensates
are included at first order. We believe that a similar pattern exists
also at higher orders and that it is possible to construct the
dictionary term by term in the perturbative expansion without changing
the identifications established at lower orders.

Our analysis is not complete: in this paper we have reproduced only a
subset of the D1D5 geometries, those that are invariant under the
$SO(4)$ rotations of the compact space $T^4$ (in the notation of
Section 2 these are the geometries with ${\cal A}_{\hat a}=0$). These
geometries are dual to those Ramond ground states of the D1D5 CFT that
are associated with the ``universal sub-sector'' of the cohomology of
$T^4$ (i.e. the $(0,0)$, $(2,2)$, $(2,0)$, $(0,2)$ forms, and the
K\"ahler $(1,1)$ form).  Hence the geometries we consider do not use
in any way the properties of the compact space $T^4$ and the results
of our computation apply, with no modification, to the case in which
the compactification manifold is K3.  We have left open the problem of
identifying which open string condensates generate the geometries
associated with the remaining even cohomology of $T^4$. We expect
these latter configurations to be related to the ones we have
considered here by the action of the supersymmetries broken by the D1
and D5 branes.  A similar approach was used in \cite{Morales:1998dz}
to study spin potentials of $\ft12$-BPS Dp-branes.  Moreover, in the
case of $T^4$, one also has the microstates associated with the odd
cohomology, which correspond, in the duality frame of the fundamental
string, to fundamental string states with pairs of fermionic
excitations \cite{Taylor:2005db}.  One can show that these states are
distinct form the ones we have considered here: indeed, all the
geometries in the class of \cite{Taylor:2005db} display $1/r^3$
corrections in the $g_{tt}$ and $g_{yy}$ metric components, which are
absent for the states we identify here. There must exist open string
condensates generating also those geometries, and it is an interesting
problem to find them.

{}From a more general point of view, we think that our computation
provides an important relation between the gravitational description
of the D1D5 microstates and their microscopic description in terms of
D-branes. Most of the studies of the fuzzball solutions focused on the
near-``horizon'' geometry and its relation with the dual CFT
description. However the full microstate geometry is asymptotically
flat and the large $r$ limit represents a regime where both the
gravitational and the D-brane descriptions are valid. Clearly, in the
string amplitude computation, we are able to explore only
perturbatively the Higgs branch, by inserting a finite number of twist
vertex operators associated to the strings stretched between the D1
and the D5 branes. However, this is sufficient to capture some of the
distinctive properties of the microstate geometries and provides a
direct support to the idea that the D-brane configurations used to
compute the entropy at $g_s=0$ evolve into fuzzballs when the string
coupling is turned on.

An advantage of our approach is that it is completely systematic and
allows, in principle, to investigate a large number of open
problems. In our opinion, two of the most important open issues are
the study of string corrections and the construction of the geometry
dual to a general 3-charge microstate. With our approach one could
derive the asymptotic expansion for such a general 3-charge geometry,
and this could provide an important clue for constructing the full
exact solution. Within our framework, one could also obtain the higher
order string corrections to the microstate geometries without the need
of guessing what term in the gravitational effective action are the
relevant one in the different configuration. We hope to come back to
these issues in a subsequent work.
 
\section*{Acknowledgements}
 We would like to thank I. Bena, S. Mathur and C. Ruef for discussions.
 This work is partially supported by the ERC Advanced Grant no.~226455, \textit{%
``Supersymmetry, Quantum Gravity and Gauge Fields''}
(\textit{SUPERFIELDS}) and by STFC under
the Rolling Grant ST/G000565/1. 
   
\providecommand{\href}[2]{#2}\begingroup\raggedright\endgroup


\begin{thebibliography}{10}

\bibitem{Sen:1995in}
A.~Sen, {\it {Extremal black holes and elementary string states}},  {\em Mod.
  Phys. Lett.} {\bf A10} (1995) 2081--2094,
  [\href{http://arxiv.org/abs/hep-th/9504147}{{\tt hep-th/9504147}}].

\bibitem{Strominger:1996sh}
A.~Strominger and C.~Vafa, {\it {Microscopic Origin of the Bekenstein-Hawking
  Entropy}},  {\em Phys. Lett.} {\bf B379} (1996) 99--104,
  [\href{http://arxiv.org/abs/hep-th/9601029}{{\tt hep-th/9601029}}].

\bibitem{Mathur:2005zp}
S.~D. Mathur, {\it {The fuzzball proposal for black holes: An elementary
  review}},  {\em Fortsch. Phys.} {\bf 53} (2005) 793--827,
  [\href{http://arxiv.org/abs/hep-th/0502050}{{\tt hep-th/0502050}}].

\bibitem{Bena:2007kg}
I.~Bena and N.~P. Warner, {\it {Black holes, black rings and their
  microstates}},  {\em Lect. Notes Phys.} {\bf 755} (2008) 1--92,
  [\href{http://arxiv.org/abs/hep-th/0701216}{{\tt hep-th/0701216}}].

\bibitem{Skenderis:2008qn}
K.~Skenderis and M.~Taylor, {\it {The fuzzball proposal for black holes}},
  {\em Phys. Rept.} {\bf 467} (2008) 117--171,
  [\href{http://arxiv.org/abs/0804.0552}{{\tt arXiv:0804.0552}}].

\bibitem{Balasubramanian:2008da}
V.~Balasubramanian, J.~de~Boer, S.~El-Showk, and I.~Messamah, {\it {Black Holes
  as Effective Geometries}},  {\em Class. Quant. Grav.} {\bf 25} (2008) 214004,
  [\href{http://arxiv.org/abs/0811.0263}{{\tt arXiv:0811.0263}}].

\bibitem{Lunin:2001fv}
O.~Lunin and S.~D. Mathur, {\it {Metric of the multiply wound rotating
  string}},  {\em Nucl. Phys.} {\bf B610} (2001) 49--76,
  [\href{http://arxiv.org/abs/hep-th/0105136}{{\tt hep-th/0105136}}].

\bibitem{Lunin:2001jy}
O.~Lunin and S.~D. Mathur, {\it {AdS/CFT duality and the black hole information
  paradox}},  {\em Nucl. Phys.} {\bf B623} (2002) 342--394,
  [\href{http://arxiv.org/abs/hep-th/0109154}{{\tt hep-th/0109154}}].

\bibitem{Lunin:2002bj}
O.~Lunin, S.~D. Mathur, and A.~Saxena, {\it {What is the gravity dual of a
  chiral primary?}},  {\em Nucl. Phys.} {\bf B655} (2003) 185--217,
  [\href{http://arxiv.org/abs/hep-th/0211292}{{\tt hep-th/0211292}}].

\bibitem{Lunin:2002iz}
O.~Lunin, J.~M. Maldacena, and L.~Maoz, {\it {Gravity solutions for the D1-D5
  system with angular momentum}},
  \href{http://arxiv.org/abs/hep-th/0212210}{{\tt hep-th/0212210}}.

\bibitem{Taylor:2005db}
M.~Taylor, {\it {General 2 charge geometries}},  {\em JHEP} {\bf 03} (2006)
  009, [\href{http://arxiv.org/abs/hep-th/0507223}{{\tt hep-th/0507223}}].

\bibitem{Kanitscheider:2007wq}
I.~Kanitscheider, K.~Skenderis, and M.~Taylor, {\it {Fuzzballs with internal
  excitations}},  {\em JHEP} {\bf 06} (2007) 056,
  [\href{http://arxiv.org/abs/0704.0690}{{\tt arXiv:0704.0690}}].

\bibitem{Lunin:2004uu}
  O.~Lunin, {\it {Adding momentum to D1-D5 system}}, {\em JHEP} {\bf 04} (2004) 054,
  [\href{http://arxiv.org/abs/hep-th/0404006}{{\tt hep-th/0404006}}].

  
\bibitem{Giusto:2004id}
S.~Giusto, S.~D. Mathur, and A.~Saxena, {\it {Dual geometries for a set of
  3-charge microstates}},  {\em Nucl. Phys.} {\bf B701} (2004) 357--379,
  [\href{http://arxiv.org/abs/hep-th/0405017}{{\tt hep-th/0405017}}].

\bibitem{Giusto:2004kj}
S.~Giusto and S.~D. Mathur, {\it {Geometry of D1-D5-P bound states}},  {\em
  Nucl. Phys.} {\bf B729} (2005) 203--220,
  [\href{http://arxiv.org/abs/hep-th/0409067}{{\tt hep-th/0409067}}].

\bibitem{Bena:2005va}
I.~Bena and N.~P. Warner, {\it {Bubbling supertubes and foaming black holes}},
  {\em Phys. Rev.} {\bf D74} (2006) 066001,
  [\href{http://arxiv.org/abs/hep-th/0505166}{{\tt hep-th/0505166}}].

\bibitem{Berglund:2005vb}
P.~Berglund, E.~G. Gimon, and T.~S. Levi, {\it {Supergravity microstates for
  BPS black holes and black rings}},  {\em JHEP} {\bf 06} (2006) 007,
  [\href{http://arxiv.org/abs/hep-th/0505167}{{\tt hep-th/0505167}}].

\bibitem{Bena:2006kb}
I.~Bena, C.-W. Wang, and N.~P. Warner, {\it {Mergers and Typical Black Hole
  Microstates}},  {\em JHEP} {\bf 11} (2006) 042,
  [\href{http://arxiv.org/abs/hep-th/0608217}{{\tt hep-th/0608217}}].

\bibitem{Bena:2008nh}
I.~Bena, N.~Bobev, C.~Ruef, and N.~P. Warner, {\it {Entropy Enhancement and
  Black Hole Microstates}},  \href{http://arxiv.org/abs/0804.4487}{{\tt
  arXiv:0804.4487}}.

\bibitem{DiVecchia:1997pr}
P.~Di~Vecchia {\em et~al.}, {\it {Classical p-branes from boundary state}},
  {\em Nucl. Phys.} {\bf B507} (1997)
  [\href{http://arxiv.org/abs/hep-th/9707068}{{\tt hep-th/9707068}}].

\bibitem{Bertolini:2000jy}
M.~Bertolini {\em et~al.}, {\it {Is a classical description of stable non-BPS
  D-branes possible?}},  {\em Nucl. Phys.} {\bf B590} (2000) 471--503,
  [\href{http://arxiv.org/abs/hep-th/0007097}{{\tt hep-th/0007097}}].

\bibitem{Dabholkar:1995nc}
A.~Dabholkar, J.~P. Gauntlett, J.~A. Harvey, and D.~Waldram, {\it {Strings as
  Solitons and Black Holes as Strings}},  {\em Nucl. Phys.} {\bf B474} (1996)
  85--121, [\href{http://arxiv.org/abs/hep-th/9511053}{{\tt hep-th/9511053}}].

\bibitem{Callan:1995hn}
C.~G. Callan, J.~M. Maldacena, and A.~W. Peet, {\it {Extremal Black Holes As
  Fundamental Strings}},  {\em Nucl. Phys.} {\bf B475} (1996) 645--678,
  [\href{http://arxiv.org/abs/hep-th/9510134}{{\tt hep-th/9510134}}].

\bibitem{Martelli:2004xq}
D.~Martelli and J.~F. Morales, {\it {Bubbling AdS(3)}},  {\em JHEP} {\bf 02}
  (2005) 048, [\href{http://arxiv.org/abs/hep-th/0412136}{{\tt
  hep-th/0412136}}].

\bibitem{Bachas:2002jg}
C.~Bachas, {\it {Relativistic string in a pulse}},  {\em Ann. Phys.} {\bf 305}
  (2003) 286--309, [\href{http://arxiv.org/abs/hep-th/0212217}{{\tt
  hep-th/0212217}}].

\bibitem{Hikida:2003bq}
Y.~Hikida, H.~Takayanagi, and T.~Takayanagi, {\it {Boundary states for D-branes
  with traveling waves}},  {\em JHEP} {\bf 04} (2003) 032,
  [\href{http://arxiv.org/abs/hep-th/0303214}{{\tt hep-th/0303214}}].

\bibitem{Billo:2002hm}
M.~Billo {\em et~al.}, {\it {Classical gauge instantons from open strings}},
  {\em JHEP} {\bf 02} (2003) 045,
  [\href{http://arxiv.org/abs/hep-th/0211250}{{\tt hep-th/0211250}}].
  
  \bibitem{Billo:2008}
M.~Billo', L.~Ferro, M.~Frau, F.~Fucito, A.~Lerda and J.~F.~Morales,
 {\it {Flux interactions on D-branes and instantons}},
  {\em JHEP}  {\bf 0810} (2008) 112,
   [\href{http://arxiv.org/abs/0807.1666}{{\tt arXiv:0807.1666}}].
  %%CITATION = JHEPA,0810,112;%%

\bibitem{Polchinski:1998rr}
J.~Polchinski, {\it {String theory. Vol. 2: Superstring theory and beyond}}, .
  Cambridge, UK: Univ. Pr. (1998) 531 p.

\bibitem{Billo:1998vr}
M.~Billo {\em et~al.}, {\it {Microscopic string analysis of the D0-D8 brane
  system and dual R-R states}},  {\em Nucl. Phys.} {\bf B526} (1998) 199--228,
  [\href{http://arxiv.org/abs/hep-th/9802088}{{\tt hep-th/9802088}}].

\bibitem{Kostelecky:1986xg}
V.~A. Kostelecky, O.~Lechtenfeld, W.~Lerche, S.~Samuel, and S.~Watamura, {\it
  {Conformal Techniques, Bosonization and Tree Level String Amplitudes}},  {\em
  Nucl. Phys.} {\bf B288} (1987) 173.

\bibitem{Morales:1998dz}
J.~F. Morales, C.~A. Scrucca, and M.~Serone, {\it {Scale independent spin
  effects in D-brane dynamics}},  {\em Nucl. Phys.} {\bf B534} (1998) 223--249,
  [\href{http://arxiv.org/abs/hep-th/9801183}{{\tt hep-th/9801183}}].

\end{thebibliography}
\end{document}